\documentclass[aps,groupedaddress,twocolumn,showpacs]{revtex4-1}
\usepackage[dvips]{graphicx}
\usepackage{dcolumn}
\usepackage{bm}
\usepackage{amssymb}
\usepackage{epstopdf}
\usepackage{amsmath}

\begin{document}

\title{Stable structural phase of potassium-doped $p$-terphenyl and its semiconducting state}

\author{Xun-Wang Yan$^{1}$}\email{yanxunwang@163.com}
\author{Zhongbing Huang$^{2}$}
\author{Miao Gao$^{3}$}\email{gaomiao@nbu.edu.cn}
\author{Chunfang Zhang$^{4}$}
\date{\today}

\affiliation{$^{1}$College of Physics and Engineering, Qufu Normal University, Qufu, Shandong 273165, China}
\affiliation{$^{2}$Faculty of Physics and Electronic Technology, Hubei University, Wuhan 430062, China}
\affiliation{$^{3}$~Faculty of Science, Ningbo University, Zhejiang 315211, China.}
\affiliation{$^{4}$College of Chemistry and Environmental Science, Hebei University, Baoding, Hebei 071002}

\begin{abstract}
The potassium-doped $p$-terphenyl compounds were synthesized in recent experiments and the superconductivity with high transition temperatures
were reported, but the atomic structure of potassium-doped $p$-terphenyl is unclear.
In this paper, we studied the structural and electronic  properties of potassium-doped $p$-terphenyl with various doping levels by the first-principles simulation.
We first find out the low energy position of K atom in intralayer interstitial space of the molecular layer, then examine whether two rows of K atoms can be accommodated in this one space, at last the effect of the interlayer arrangement between adjacent two molecular layers on total energy are taken into account.
Our results show that the doped K atoms prefer to stay at the bridge site of single C-C bond connected two phenyls instead of locating at the site above the phenyl ring, distinct from the situation of K-doped picene and phenanthrene.
Among the possible structural phases of K$_x$-$p$-terphenyl,
the K$_2$-$p$-terphenyl phase with P2$_1$2$_1$2$_1$ group symmetry is determined to be most appropriate, which is different from the one in recent report.
The stable K$_2$-$p$-terphenyl phase is semiconducting with a energy gap of 0.3 eV and the bands from the lowest unoccupied molecular orbitals are just fully filled by the electrons transferred from K atoms.
\end{abstract}

\pacs{74.70.Kn, 74.20.Pq, 61.66.Hq, 61.48.-c}

\maketitle
\section{Introduction}
Recently, the intercalation of alkali metal into polycyclic aromatic hydrocarbons to synthesize the new electronic materials  has been an intriguing topic in condensed matter physics and material science fields.
In 2010, potassium-doped picene (C$_{22}$H$_{14}$) was found to have a superconducting phase with the transition temperature (T$_c$) of 18 Kelvin.\cite{Mitsuhashi2010} Since then, other aromatic hydrocarbons, such as  phenanthrene, coronene, dibenzopentacene, chrysene, pentacene, tetracene and anthracene
were adopted to synthesize the samples for exploring the superconductivity. \cite{Kubozono2016}
These molecular solids are all condensed aromatic compounds with some common features.
In a molecule the benzene rings are fused together along a zigzag or straight line, and
in the crystal the molecules are arranged in a herringbone pattern to form a molecular layer;
the interstitial space in molecular layer is large enough to accommodate the metal dopant.

The detailed atomic structure of metal-doped aromatics is the key factor to understand the electronic and magnetic properties, but it is difficult to be determined in experiments because of their degradation in air and the low sample quality. \cite{Wang2012, Artioli2014, Kubozono2016, Mitsuhashi2010, Kubozono2011, Xue2012}
The theoretical simulation is one of the important research approaches for exploring the structural and electronic properties of these new metal-doped aromatic hydrocarbons, and lots of significant results have been obtained. \cite{Kosugi2011, DeAndres2011a, DeAndres2011, Kubozono2011, Giovannetti2011, Huang2012, Ruff2013, Naghavi2013a, Naghavi2014, Yan2013, Yan2014, Yan2016a,Yan2016, Wang2017c}
For example, the dopants are intercalated in the intralayer space instead of interlayer space in molecular crystal;
the electrons are transferred from the doped metal to molecule and the charge is delocalized over the the whole organic crystal by the $\pi$ molecular orbitals;
the energy band group from the lowest unoccupied molecular orbital (LUMO) or LUMO + 1 plays an important role in electronic behaviour around Fermi energy.
These electronic properties are the vital prerequisites for us to understand the mechanism of step-like magnetization transition possibly related to superconductivity in metal-doped aromatic hydrocarbons.

$p$-terphenyl crystal is another aromatic hydrocarbon different from picene and phenanthrene crystals, in which the phenyl rings of molecule are connected by single C-C bond rather than fused together. Recently, potassium-doped $p$-terphenyl (PTP) was reported to have a surprising superconducting phases with the transition temperatures up to 120 Kalvin. \cite{Wang2017a}
Soon after that, the measured photoemission spectra of K-doped terphenyl showed that a low energy gap can persist to 120 K which most likely originated from electron pairing.\cite{Li2017}
The temperature dependence of magnetic susceptibility measured for K-doped terphenyl and quaterphenyl displayed the step-like transition at about 125 K.\cite{Liu2017}
The discoveries have great scientific significance because the transition temperature of 123 Kalvin is so high.
But, due to the low quality samples of K-doped terphenyl, the detailed crystal structures and the stoichiometric compositions are not clear in experiment.
A recent theoretical work reported by Prof. Zhong proposes that K$_x$PTP at $x$ = 2.5 is most stable.\cite{Zhong2018}
Different from Zhong's conclusion, we emphasize that K$_2$PTP is the most typical product for K-doped terphenyl compounds.
We have performed the structural and electronic properties calculations of K-doped terphenyl.
Our results demonstrate that the stoichiometry ratio of K and molecule in K-doped PTP compounds is 2:1, which accords with the bipolaron occurrence based on the Raman spectra analysis in experiment.\cite{Wang2017a}
The K$_2$-F phase of K$_2$PTP with P2$_1$2$_1$2$_1$ group symmetry is found to be the most appropriate configuration,
the ground state of which is semiconducting with a energy band gap of 0.3 eV.
Besides, it is noteworthy that the K$_2$-F phase is similar to Cs$_2$phenanthrene in molecular layer arrangement and crystalline symmetry, measured in a recent experiment.\cite{Romero2017}

\section{Computational details}
In our calculations, the plane wave basis sets and pseudopotential method were used. The generalized gradient approximation
(GGA) with Perdew-Burke-Ernzerhof (PBE) formula was adopted for the exchange-correlation potentials. \cite{PhysRevLett.77.3865}
The projector augmented-wave (PAW) pseudopotential for C, H and K elements are from the subfolder C$\_s$, H$\_s$ and K$\_{sv}$ in the pseudopotential package potpaw$\_$PBE.52 supplied by Vienna Ab initio simulation package (VASP) website. \cite{PhysRevB.47.558}
A mesh of $4\times 4\times 4$ k-points was used for the relaxation of the lattice parameters and the internal atomic positions, and $8\times 8\times 6$ k-point mesh was used for the DOS calculations.
The plane wave basis cutoff is 500 eV, and the convergence thresholds of the total energy, force on atom and pressure on cell are 10$^{-5}$ eV, 0.005 eV/\AA ~and 0.1 KBar respectively.
The van der Waals (vdW) interaction is included in our calculations,
and the van de Waals functional we used is the vdW-DF2 scheme, proposed by Langreth and Lundqvist {\it et al}.
\cite{PhysRevLett.92.246401}

\section{Results and Analysis}
PTP crystalizes in the space group P2$_1$/a.
In pristine crystal, the PTP molecules form a herringbone structure in a molecular layer, and the molecule layers are stacked along the $c$ axis direction.
In order to inspect the applicability of  the C and H pseudopotentials
 and the parameters selected in our calculations,
we first simulated the crystal structure of pristine $p$-terphenyl.
The optimized lattice parameters ($a = 8.106$ \AA, $b = 5.613$ \AA, $c =13.613$ \AA ~and $\beta =92.02  ^{\circ}$) have a good agreement with the experimental ones ($a = 7.945 $ \AA, $b = 5.581 $ \AA, $c = 13.628 $ \AA ~and $\beta = 92.73 ^{\circ}$).
 The consistency is a reliable basis for further exploring the structure of K-doped PTP.

\subsection{Analysis of K dopant positions in K$_x$PTP}
In a molecular layer, the molecule arrangement with a herringbone pattern leaves large interstitial spaces among molecules. One space is enclosed by four neighbor molecules and named as a hole in this paper, shown in Fig. \ref{structure-1}.
The metal atoms can be inserted into these holes and interact with the molecules to stabilize the doped molecular crystal.
More specifically, the K atom positions in the hole can sit above the carbon ring or close to the single C-C bond between two rings of terphenyl molecule.
The length of an interstitial hole is about 14 \AA, and the K-K bond length in bulk potassium metal is 4.54 \AA, so the maximum number of K atoms aligned in a line along the hole is three. Or else, the repulsion between K atoms will decrease the stability of K$_x$terphenyl system.
Under such constraint, the doping schemes become simple.
One unit cell of pristine terphenyl has two holes averagely.
 Usually, K atoms are uniformly distributed in each of two holes. Another case is that one of two hole is filled by K atom and the other is empty, which is mentioned in Ref. \citenum{Zhong2018}.

The interlayer space between two adjacent molecular layers is another possible region to accommodate the dopants.
However, previous studies have proved that dopants do not favor the interlayer positions,\cite{Kosugi2011, Yan2016}
because the metal dopants are repelled by the H atoms at the end of PTP molecules.
Hence, in what follows, K atoms are placed in the intralayer holes with various configurations to build the initial structures, then the atom positions and lattice parameters are relaxed to reach the optimized ones.
This procedure is widely adopted in previous works to explore the structures of metal intercalated polycyclic aromatic hydrocarbons.\cite{Kosugi2011,DeAndres2011a,DeAndres2011,Yan2013}

In addition, the arrangement pattern of molecular layers can affect the interaction between two adjacent layers in K-doped terphenyl compounds. Inspired by the structures of K$_2$picene and K$_2$pentacene reported in experiment,\cite{Romero2017} we build two structural phases of K$_2$PTP with P2$_1$2$_1$2$_1$ and P2$_1$/c group symmetry to examine the influence of interlayer alignment on the system stability.

\subsection{Crystal structure of K$_x$PTP with K dopants in intralayer interstitial space}
Although the K metal and pristine PTP are mixed in the stoichiometry ratio of 3:1 in experiment,
the ratio of K-doped terphenyl is not known because of the KH generation in experiment.\cite{Zhong2018, Romero2017}
In fact, the real chemical composition and the exact structure of K-doped aromatic hydrocarbons are always the bottleneck problem in the research field.
In order to explore the stable structure and the reasonable ratio of K atom and terphenyl molecule, several $x$ values are adopted to perform the calculations of K$_x$PTP compounds.

\subsubsection{K$_1$PTP}

\begin{figure}.
\begin{center}
\includegraphics[width=7.5cm]{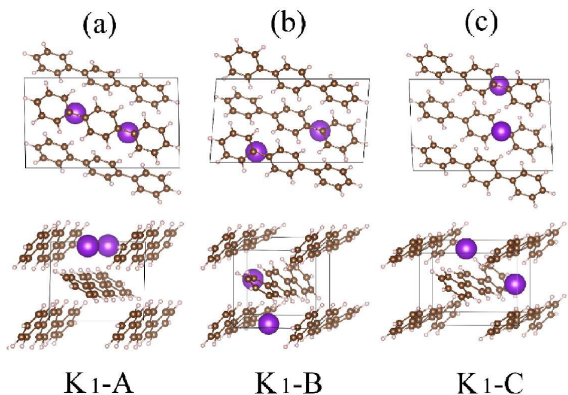}
\caption{
 The crystal structures of three structural phases for K$_1$PTP. The side views along $b$ axis (top panel) and the front views along $c$ axis (bottom panel) are shown in (a), (b) and (c) for the unit cells of K$_1$-A, K$_1$-B and K$_1$-C, respectively.  Their unit cells are all composed of two molecules and two K atoms. Large spheres represent K atoms. One interstitial space accommodating K atoms is named as a hole.
  } \label{structure-1}
\end{center}
\end{figure}

The initial structure of K$_x$PTP is based on the pristine PTP structure, which unit cell contains two holes.
Two K atoms are inserted into the two holes with various initial positions to form the crystal cell of K$_1$PTP, and
after full relaxation three optimized structures are obtained, named as K$_1$-A, K$_1$-B and K$_1$-C shown in Fig.\ref{structure-1}.
The lattice parameters for three structural phases of K$_1$PTP, K$_1$-A, K$_1$-B and K$_1$-C, are listed in Table \ref{opti-latt}.
Among three K$_1$PTP phases, K$_1$-A has the lowest energy, which is 0.18 eV and 0.70 eV lower than the total energies of K$_1$-B and K$_1$-C.
Therefore, for K$_1$PTP, K$_1$-A is the most stable structure with two K atoms in the same hole,
which configuration is similar to the La-doped phenanthrene reported in previous works. \cite{Naghavi2013,Yan2013}

\subsubsection{K$_2$PTP}
\begin{figure}.
\begin{center}
\includegraphics[width=7.50cm]{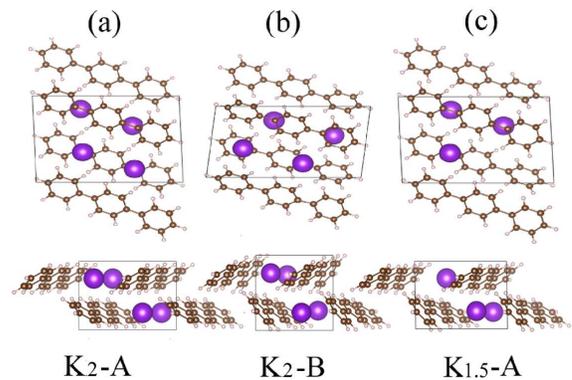}
\caption{
  Crystal structures of K2-A and K$_2$-B phases for K$_2$PHA and K$_{1.5}$-A phase for K$_{1.5}$PHA. (a), K$_2$-A phase;  (b), K$_2$-B phase;  (c), K$_{1.5}$-A phase . Large spheres represent K atoms.} \label{structure-2}
\end{center}
\end{figure}

By adding two K atoms into the unit cell of the K$_1$PTP phases and doing the full relaxations,
two structural phases of K$_2$PTP are obtained, shown in Fig. \ref{structure-2}(a) and (b).
Four K atoms in K$_2$-A phase are all near the single C-C bonds, but in K$_2$-B phase there are two K dopants close to phenyl rings at the end of PTP molecule.
The K$_2$-A energy per unit cell is 0.04 eV lower than the K$_2$-B energy,
thus the K$_2$-A phase is more reasonable than K$_2$-B.
The structural character of K$_2$-A is that four K atoms in a cell happen to fill in four sites near the single C-C bonds.
We also examine the K doping effect on the length of single C-C bond between two phenyl rings of PTP molecule,
the single C-C bond length in  K-doped PTP crystal is reduced from 1.495 \AA~ in pristine crystal to 1.438 \AA.

\begin{table}
  \caption{The lattice parameters for various phases of K$_x$PHA with x= 1, 1.5, 2, 2.5, 3.}
  \label{opti-latt}
  \begin{tabular}{lllllll}
    \hline
          phases    & a  & b & c &$\alpha$ &$\beta$ & $\gamma$ \\
    \hline
    K$_1$-A   & 7.76 & 6.48 & 13.22 & 89.3 & 90.9 & 88.8 \\
    K$_1$-B   & 7.33 & 6.82 & 13.40 & 90.0 & 85.7 & 90.0  \\
    K$_1$-C   & 8.49 & 6.03 & 12.97 & 90.0 & 92.5 & 90.0 \\
           &     &   &     &  &   &   \\
    K$_2$-A   & 9.09 & 6.31 & 12.65 & 90.0 & 93.7 & 90.0  \\
    K$_2$-B   & 7.39 & 7.14 & 13.57 & 90.0 & 83.5 & 90.0  \\
    K$_2$-C   & 9.13 & 6.17 & 13.40 & 83.3 & 91.1 & 94.8  \\
    K$_2$-D   & 10.15 & 10.75 & 13.90 & 90.0 & 97.3 & 90.0  \\
    K$_2$-E   & 9.00 & 6.33 & 25.41 & 90.0 & 92.7 & 90.0  \\
    K$_2$-F   & 9.19 & 6.37 & 24.36 & 90.0 & 90 & 90.0  \\
    K$_{1.5}$-A   & 8.60 & 6.37 & 12.85 & 90.6 & 92.3 & 90.9  \\
        &     &   &     &  &   &   \\
    K$_3$-A   & 7.44 & 7.48 & 14.42 & 90.0 & 76.7 & 90.0  \\
    K$_3$-B   & 7.07 & 7.09 & 18.88 & 90.0 & 77.6 & 90.0  \\
    K$_{2.5}$-A   & 7.26 & 7.28 & 14.44 & 89.9 & 76.8 & 90.1  \\
    K$_{2.5}$-B   & 9.06 & 6.28 & 14.61 & 83.2 & 106.5 & 99.3  \\
    \hline
  \end{tabular}
\end{table}

\subsubsection{K$_3$PTP}
\begin{figure}.
\begin{center}
\includegraphics[width=7.50cm]{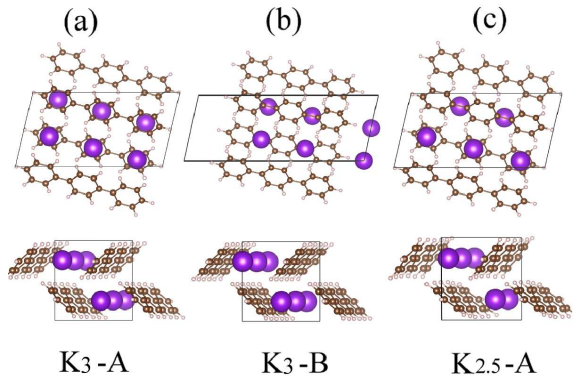}
\caption{
  Crystal structures of K$_3$-A and K$_3$-B phases for K$_3$PTP and K$_{2.5}$-A phase for K$_{2.5}$-APHA. (a), K$_3$-A phase with all six K atoms in the intralayer space. (b), K$_3$-B phase with four K atoms in the intralayer space and two in the interlayer space.  (c), K$_{2.5}$-A phase . Large spheres represent K atoms.} \label{structure-3}
\end{center}
\end{figure}

For K$_3$PTP, we present two optimized configurations, marked as K$_3$-A and K$_3$-B phases shown in Fig. \ref{structure-3}(a) and  (b).
 The optimized lattice parameters for these two structural phases are listed in Table \ref{opti-latt}.
In the unit cell of K$_3$-A phase, six K atoms are uniformly distributed in the intralayer region. Because the repulsion among K atoms, the K atoms are close to phenyl rings, instead of single C-C bonds.
K$_3$-B structure is derived from K$_2$-A by adding two K atoms in interlayer region in the elongated cell.
It is noted that the energy of K$_3$-B is 0.50 eV higher per unit cell with respect to that of K$_3$-A, which confirms the previous conclusion once again that metal dopants do not favor the interlayer space in aromatic molecular solid. \cite{Kosugi2011, Yan2016}

\subsubsection{K$_{1.5}$PTP and K$_{2.5}$PTP}
For K-doped PTP, the K dopant concentration may deviate from the integer ratio of K atom to molecule, such as 2 : 1 and 3 : 1.
We present the structures of K$_{1.5}$PTP and K$_{2.5}$PTP in Fig. \ref{structure-2}(c) and Fig. \ref{structure-3}(c), marked as K$_{1.5}$-A and K$_{2.5}$-A.
 The optimized lattice parameters for these two structural phases are listed in Table \ref{opti-latt}.
The lattice parameters of the K$_{1.5}$-A unit cell are close to the ones of K$_2$-A, so K$_{1.5}$-A can be considered as the K$_2$-A phase with a K atom defect.
Likewise, K$_{2.5}$-A phase can be regarded as the K$_3$-A phase with a K atom defect due to its similar lattice parameters to the K$_3$-A phase.

In a brief summary, several K$_x$PTP structural phases are built by inserting K dopant into the intralayer interstitial space.
When one hole accommodates one or two K atoms in K$_1$TPT, K$_2$TPT and K$_{1.5}$TPT structures, K atoms are relaxed to the sites near the single C-C bonds, which coordinate is about $\frac{1}{3}c$ or $\frac{2}{3}c$.
Since the four single C-C bonds are around the site instead of carbon rings,
the specific site in the intralayer interstitial space has the maximum room. The K atoms at the sites would have the minimum repulsion from C and H atoms of terphenyl molecule.
When three K atoms are placed into one hole in K$_3$TPT and K$_{2.5}$TPT structures, K atoms sit at both ends and midpoint of the hole, namely close to carbon rings of terphenyl molecule, which results from the repulsion between K atoms.
Therefore, the site near the single C-C bond is the most favorable site for K dopant in the molecular layer.

\subsection{Crystal structure of K$_{x}$PTP with two rows of K atoms filling in one intralayer interstitial space}
There exists another kind of K atom distribution in the intralayer space, i.e. one hole is filled by two rows of K atoms while the neighbor hole keeps empty for the two holes in a unit cell. On the base of the pristine terphenyl cell, we place four K atoms into one of two holes to form the K$_2$PTP unit cell, and the structural phase is named as K$_2$-C.
Cs-doped phenanthrene compounds were synthesized and the detailed structures were determined in recent experiment.\cite{Takabayashi2017} Inspired by the Cs$_2$phenanthrene structure with P2$_1$/a group symmetry, we imitate the structure and build the K$_2$-D phase of K-doped terphenyl, which unit cell contains four molecules and eight K atoms.
A recent theoretical work points out that the interstitial hole can accommodate five K atoms to form a kind of K$_{2.5}$PTP structure.\cite{Zhong2018}
According to the structure in Fig. 2 (e) in Ref. \citenum{Zhong2018}, we rebuild the K$_{2.5}$PTP structure and call it K$_{2.5}$-B phase.

\begin{figure}.
\begin{center}
\includegraphics[width=7.5cm]{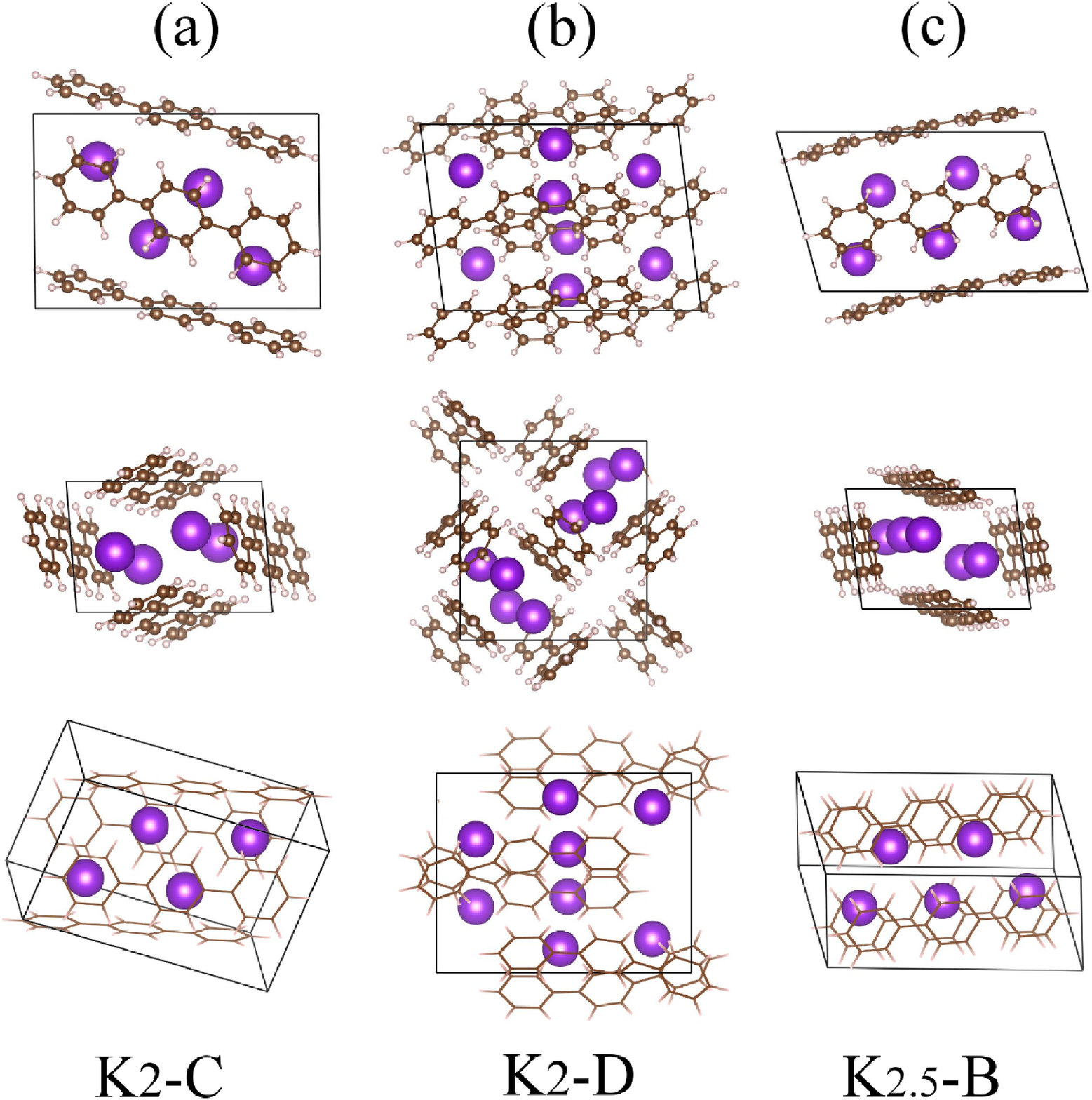}
\caption{
  Three structural phases of K$_2$PTP and K$_{2.5}$PTP with one intralayer hole accommodating two rows of K atoms. (a), K$_2$-C phase,
  (b), K$_2$-D phase,
  (c), K$_{2.5}$-B phase.
  The upper and middle panels is the side views along $b$ and $c$ axes, and the bottom panel is to show the K atom positions in the interstitial hole in the molecular layer. } \label{structure-tworowK}
\end{center}
\end{figure}

K$_2$-C, K$_2$-D and K$_{2.5}$-B structures are displayed in Fig. \ref{structure-tworowK}. The upper and middle panels are the side views along $b$ and $c$ axes, and the bottom panel shows the K atom positions in the interstitial hole in the molecular layer.
 As shown in Fig. \ref{structure-tworowK}(a) bottom panel, for the up and down molecules in the K$_2$-C cell, four K atoms are close to the single C-C bonds; for the front and back molecules, the K atoms are close to the carbon rings.
In Fig. \ref{structure-tworowK}(b) bottom panel, for the K$_2$-D unit cell, all four K atoms locate at the sites near the single C-C bonds but some of molecules have an obvious shift in $c$ axis direction.
 In K$_{2.5}$-B phase, upper row of K atoms are close to single C-C bonds of two upper molecules, and bottom row of K atoms are close to three carbon rings of two bottom molecules, see Fig. \ref{structure-tworowK}(c) bottom panel.

When a interstitial hole is filled with two rows of K atoms, some of them are pushed to the sites near carbon ring because of the repulsion among these K atoms, leading to the system energy going up with respect to K$_2$-A phase. Thus, the two rows of K atom in a hole is not a energy favorable configuration.
Among above four K$_2$PTP phases, i.e. K$_2$-A, K$_2$-B,  K$_2$-C, and K$_2$-D, the energy of K$_2$-A is lower than other three phase energies. Here, we regard the K$_2$-A phase as the typical structure for all K-doped terphenyl compounds, in which all low energy sites are just fully occupied and K dopants have the most uniform distribution in molecular layer.

\subsection{Crystal structure of K$_{2}$PTP with the unit cell consisting of two molecular layers }

 In previous theoretical calculations, the unit cell of metal-doped polycyclic aromatic hydrocarbons is limited to one molecular layer for the reason of computational time economization. So the different arrangement of molecular layers has rarely been involved in the theoretical works reported before. But for K-doped picene and pentacene in experiment, the measured unit cells of K$_2$picene and K$_2$pentacene are all composed of two layers of molecules, which demonstrates that the interlayer stacking pattern is of importance for this class of materials.
 According to the space group symmetry and molecular orientation of K$_2$picene and K$_2$pentacene in experiment,
 we build two kinds of structural phases for K-doped terphenyl, named as K$_2$-E and K$_2$-F, whose unit cells include two molecular layers.
 It should be emphasized that for the two structures, their atomic configuration of molecular layer is similar to that of the K$_2$-A phase. Specifically, the K positions are all in the low energy sites near single C-C bonds.

\begin{figure}.
\begin{center}
\includegraphics[width=7.5cm]{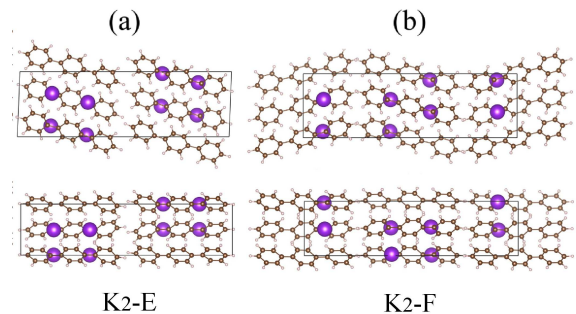}
\caption{
  Two structural phases of K$_2$PTP with the unit cells containing two molecular layers. (a), The side views of K$_2$-E phase with P2$_1$/c group symmetry along $b$ axis (upper) and $a$ axis (lower). (b), The side views of K$_2$-E phase with P2$_1$2$_1$2$_1$ group symmetry along $b$ axis (upper) and $a$ axis (lower).  The molecular orientations in two phases are distinct.   } \label{structure-p21c-p212121}
\end{center}
\end{figure}

 Our aim is to consider the influence of stacking pattern of molecular layers on the total energy of K$_2$PTP.
 We hope to keep the intralayer atomic configuration of K$_2$-A phase and only change the interlayer arrangement to build the new structure.
 The expectation is achieved by K$_2$-E and K$_2$-F structures.
 Fig. \ref{structure-p21c-p212121}(a) and (b) show the atomic structures of K$_2$-E and K$_2$-F with P2$_1$/c and P2$_1$2$_1$2$_1$ group symmetry, respectively.
 By comparing K$_2$-A, K$_2$-E and K$_2$-F phases, we find that their symmetry operations are all related to screw operation: screw around $b$ then translate along $a$ for K$_2$-A, screw around $b$ then translate along $c$ for K$_2$-E and screw around each of three crystal axes for K$_2$-F.
 Different interface configuration between two adjacent molecular layers is related to their symmetry operations, and result in the difference of interlayer interaction.

 We should pay special attention to K$_2$-F phase. The molecular orientation in adjacent two layers is obviously deflected from one to another, which leads to a more close packing of molecular layers than K$_2$-E and K$_2$-A. The volume of K$_2$-F unit cell decrease 1.2\%-1.5\% with respect to K$_2$-E and K$_2$-A phases in our calculations. Meanwhile, the energy per unit cell of K$_2$-F phase is 0.237 eV and 0.226 eV lower than the energies of K$_2$-E and K$_2$-A.
 So far, as a stable structure of K$_x$PTP, K$_2$-F structure is derived after considering various factors, including the intralayer K dopant position, occupation number and interlayer arrangement.
 The crystal structure data of K$_2$-F phase is given in the Supplemental Material\cite{suppl} in cif format.

\subsection{Formation energy of K$_x$PTP with $x$ = 1, 1.5, 2, 2.5, 3}
\begin{figure}.
\begin{center}
\includegraphics[width=7.50cm]{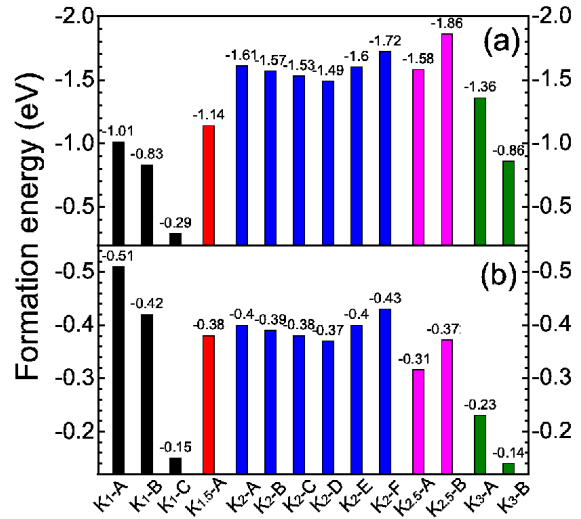}
\caption{
  (a), The formation energies of various structural phases for K$_x$PTP, the values is the half of formation energies for K$_2$-D, K$_2$-E and K$_2$-F because their unit cells are twice the unit cells of other phases. (b), The average formation energy per K atom for various structural phases.  } \label{formation-4}
\end{center}
\end{figure}

The alkali metal K is mixed with polycyclic aromatic hydrocarbon $p$-terphenyl to synthesize the new K$_x$PTP compounds.
The appropriate stoichiometry ratio and the stability of various structural phases for K$_x$PTP compounds can be reflected by their formation energies, and the formula of formation energy are shown as follows.

\begin{eqnarray}
 \begin{aligned}
  E_{\rm f} =E_{\rm K_xPTP}-E_{\rm PTP}-n \cdot E_{\rm K}
 \end{aligned}
\end{eqnarray}

where $E_{\rm K_xPTP}$, $E_{\rm PTP}$, and $E_{\rm K}$ are the energies of K$_x$PTP, pristine PTP, and  one K atom in bulk metal, respectively.
The calculated formation energies for various structural phases are displayed in Fig. \ref{formation-4}.
For the sake of contrast, the formation energy values of K$_2$-D, K$_2$-E and K$_2$-F phases are divided by 2 because their unit cells are twice of the unit cell of other structural phases.
From Fig. \ref{formation-4}(a) we can see that the formation energy increases with $x$ from 1 to 2.5, and then decreases at $x$ = 3.
K$_{2.5}$-B and K$_2$-F have the maximum energy of -1.86 eV and second maximum energy of -1.72 eV (minus sign is energy release).
We focus on the low energy structural phases at the special doping levels, i.e. K$_1$-A,  K$_{1.5}$-A, K$_2$-F, K$_{2.5}$-B and K$_3$-A phases.
When the structural phases change from K$_1$-A to K$_{1.5}$-A, K$_2$-F, K$_{2.5}$-B and K$_3$-A, the formation energies do not change uniformly with the doping level increasing. The values below are how much the formation energy is increased when one more K atom is added into K$_x$PTP unit cell.

\begin{eqnarray}
  E_{\rm f}(K_{1.5}-A)-E_{\rm f}(K_1-A)-E_{\rm K} = -0.13 ~eV, \label{eq.2} \\
  E_{\rm f}(K_2-F)-E_{\rm f}(K_{1.5}-A)-1 \cdot E_{\rm K} = -0.58 ~eV, \label{eq.3} \\
  E_{\rm f}(K_{2.5}-B)-E_{\rm f}(K_2-F)-1 \cdot E_{\rm K} = -0.14 ~eV, \label{eq.4} \\
  E_{\rm f}(K_3-A)-E_{\rm f}(K_{2.5}-B)-1 \cdot E_{\rm K} = +0.50 ~eV,  \label{eq.5}
\end{eqnarray}

The value of +0.5 eV in expression (\ref{eq.5}) means that the K doping level of $x$ = 3 is difficult to be achieved in K$_x$PTP,
because the process of adding K atoms into the K$_{2.5}$-B unit cell to form the K$_3$-A phase will lift a large energy.
Meanwhile, the generation of K$_2$-F phase is most easy from K$_{1.5}$-A phase due to the energy decline of -0.58 eV in expression (\ref{eq.3}).

The average formation energy per K dopant is another important quantity to reflect the stability of doped compounds,
which is derived by dividing the total formation energy by the number of K atoms, shown in Fig. \ref{formation-4}(b). The maximum value of -0.51 eV belongs to K$_1$-A phase and the second maximum energy of -0.43 eV belongs to K$_2$-F.
For K$_2$-A and K$_2$-E phases, the values are both -0.40 eV.
The average formation energy per K atom is -0.37 eV for K$_{2.5}$-B phase, and the value is -0.23 eV for K$_3$PTP phases.
For K$_2$-A, K$_2$-E, K$_2$-F and K$_1$-A phases, they share one structural feature in common that one intralayer hole contains two K atoms siting in C-C single bond positions. The average formation energies for four phases have the larger values than other phases, which provide hard evidence that the dopant site close to C-C single bond is preferred in energy.
Among these structural phases, K$_2$-F should be emphasized because it is a typical structural phase, which relates to its large formation energy, low energy site of K dopant, and appropriate dopant occupation number.

\subsection{Electronic structure of K$_x$PTP with K$_2$-F phase}
\begin{figure}
\begin{center}
\includegraphics[width=7.50cm]{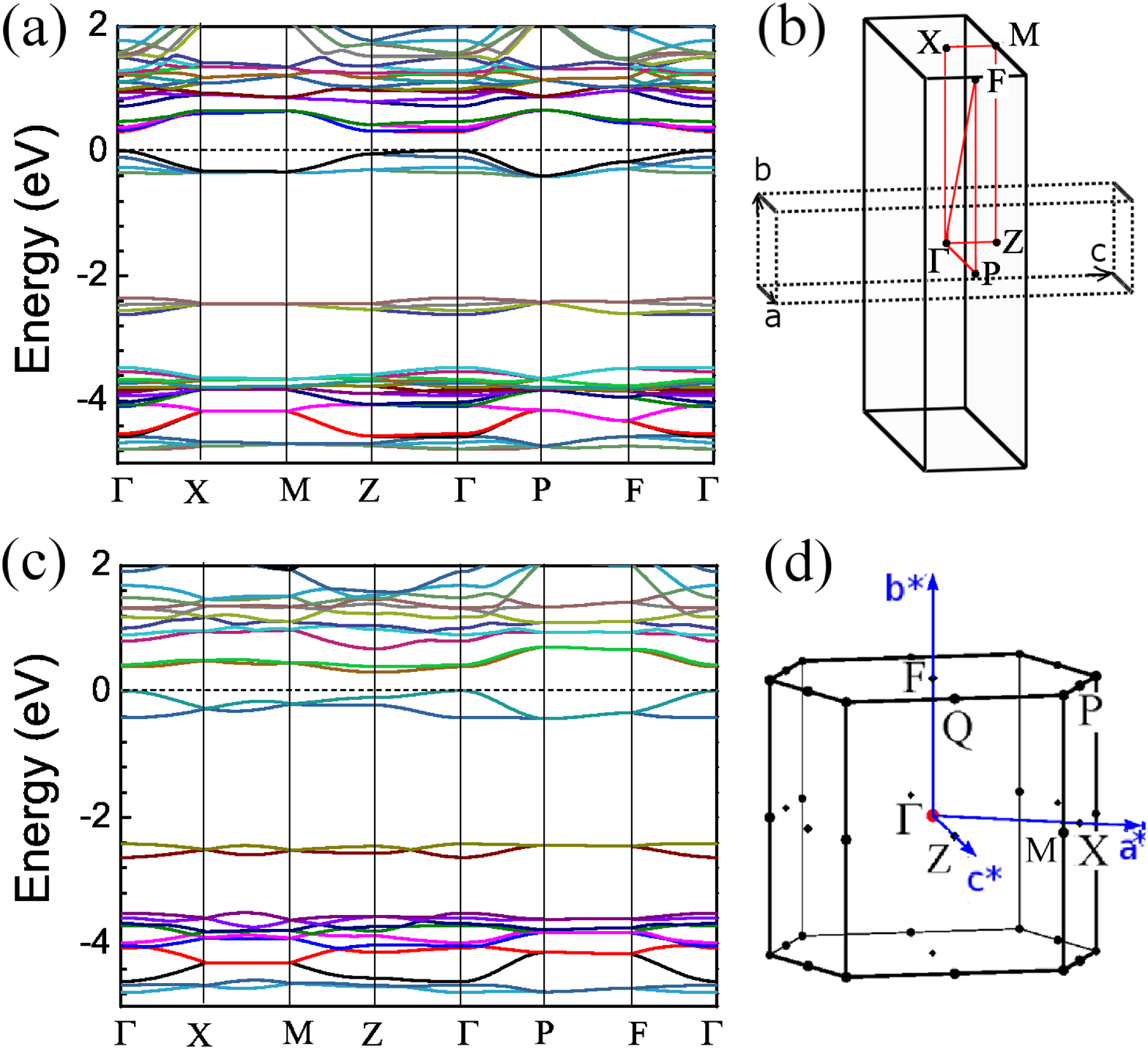}
\caption{
  (a) and (b), Energy band structure of K$_2$PTP with K$_2$-F phase and the Brillouin Zone with high symmetry points, and the unit cell is also plotted with the dashed line box. (c) and (d), Energy band structure of K$_2$PTP with K$_2$-A phase and the Brillouin Zone with high symmetry points. Fermi level is set to zero.} \label{K2A-band}
\end{center}
\end{figure}

The band structure of K$_2$-F phase for K-doped terphenyl is displayed in Fig. \ref{K2A-band}(a), and the reciprocal Brillouin zone is shown in Fig. \ref{K2A-band}(b).
At first, there is a energy gap of $\sim$0.3 eV at the Fermi energy, indicating that the K$_2$-F phase is a semiconductor. The feature is similar to the semiconducting K$_2$picene with a gap of 0.1 eV. \cite{Kosugi2011, Kusakabe2012}
Four bands below Fermi energy, which arise from the LUMO orbitals of four molecules in one unit cell, are all filled by electrons from eight K atoms. Consequently, a energy gap near Fermi energy is opened
to lead to a semiconducting state.
Also, there exists another large energy gap in the scope of -2.4 $\sim$ -0.4 eV, which is relevant to the energy interval between the highest occupied orbital (HUMO) and LUMO.

Secondly, on a whole, these bands are flat and have small dispersion in energy.  This is because the interactions among molecules are weak in molecular crystal. In Fig. \ref{K2A-band}(b), the crystal cell is marked by dash line. $a$ and $b$ direction correspond to $\Gamma$-X and $\Gamma$-P in reciprocal space and the $c$ axis relates to $\Gamma$-Z and X-M direction. The bands along $\Gamma$-Z and X-M are more flat than the bands along $\Gamma$-X and $\Gamma$-P, which reflect that the interaction between two molecular layer is more weaker than the interaction in molecular layer.

Thirdly, Fig. \ref{K2A-band}(c) and (d) display the energy bands and the Brillouin zone of K$_2$-A unit cell.
We compare the band structures of K$_2$-A and K$_2$-F phases and find that they have a nearly identical distribution,
including the band shape and the interval among bands.
The similarity of band structure demonstrate that the electronic properties of K-doped terphenyl are determined primarily by the atomic configuration of molecular layer and the interlayer stacking pattern has a small influence on them.
Therefore, the band structure of K$_2$-A and K$_2$-F are representative for K-doped terphenyl compounds.

\begin{figure}
\begin{center}
\includegraphics[width=7.5cm]{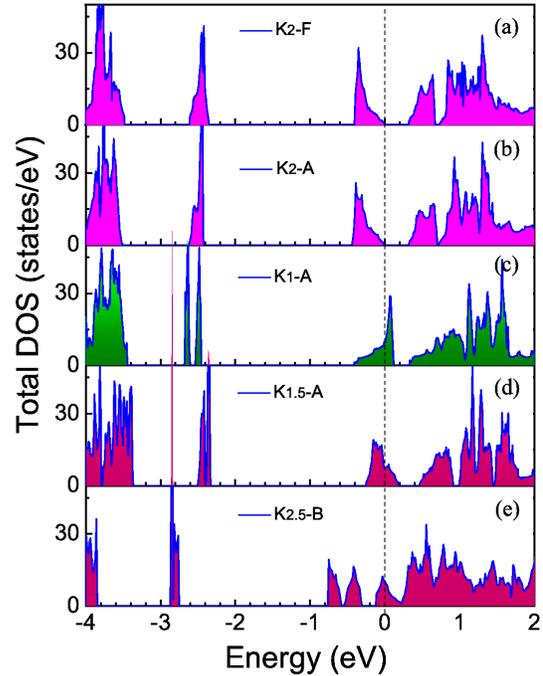}
\caption{
   Total DOS of K$_2$-F, K$_2$-A, K$_1$-A, K$_{1.5}$-A and K$_{2.5}$-B phases. Fermi level is set to zero and is marked by the vertical dash line.} \label{K1-K2-A-dos3}
\end{center}
\end{figure}

The total density of states (DOS) of K-doped PTP with several structural phases are shown in Fig. \ref{K1-K2-A-dos3}.
 We first inspect the DOS of K$_2$-F phase in Fig. \ref{K1-K2-A-dos3}(a), the peaks centered at -2.5 eV and -0.3 eV are related to HOMO and LUMO of PTP molecule.
 With the ratio of 2:1 between K and molecule, the states from LUMO orbitals are just fully filled by the electrons donated by K atoms.
  The energy gap of 0.3 eV above the Fermi energy is associated to the occupied bands from LUMO states and the empty bands from LUMO+1 states.
 As expected, the DOS peaks distribution of K$_2$-A phase in Fig. \ref{K1-K2-A-dos3}(b) is very similar to that of K$_2$-F phase because their intralayer atomic configuration are alike.
 When the ratio of K and molecule is less or greater than 2 : 1, the Fermi energy of K$_2$PTP will be lowered or raised with respect to the energy of K$_2$PTP.
 For K$_{1}$-A and K$_{1.5}$-A phase, the electronic states from molecule LUMO orbitals are half-filled and the systems are metallic, as shown in Fig. \ref{K1-K2-A-dos3}(c) and (d), where the DOS values at Fermi energy are 9.5 and 7.3 states/eV per unit cell for K$_{1}$-A and K$_{1.5}$-A phases.
For K$_{2.5}$-B phase, the fermi energy is located in the energy scope of LUMO+1 states and the states from LUMO orbitals are fully filled. The DOS value at Fermi level is 10.1 states/eV per unit cell, see Fig. \ref{K1-K2-A-dos3}(e).

\section{Conclusions and Discussions}

In summary, we have investigated the effect of various factors on the crystal structure of K-doped terphenyl, including the stoichiometry ratio of dopant and molecule, the K doping position, the number of dopants in one intralayer interstitial hole and the interlayer stacking pattern.
The results indicate that the sites near the single C-C bond in the interstitial hole are the most favorable positions in energy.
Two K atoms in such a hole is found to be the optimized occupation, while the situation of two rows of K atoms in one hole is not energetically favorable.
 The interlayer arrangement with P2$_1$2$_1$2$_1$ group symmetry can effectively enhance the stability of the system.
Consequently, the K$_2$PTP with K$_2$-F phase is deduced to be the most appropriate structure.
Besides, the electronic structure calculations show K$_2$-F phase is a semiconductor with a small energy gap of 0.3 eV.

In spite of the preferred K$_2$-F phase, there exist several structural phases of K$_x$PTP at $x$ = 2, which have the small energy differences to K$_2$-F phase. Therefore, such meta-stable phases have the some possibility to occur in samples at certain temperatures.
When the nominal ratio of K and molecule is less than or equal to 1:1, K$_1$-A phase is more favorable.
When the ratio is greater than 2 : 1 in experiment, K$_{2.5}$-B phase can be generated.
The synthesis of K$_3$PTP phase is difficult because it will decompose to K$_2$-F phase and K bulk metal in viewpoint of formation energy.
The existence of multiple structures for K-doped $p$-terphenyl is supported by our calculations, which is in agreement with the reported experiments about alkali-metal-doped aromatic hydrocarbons.
The superconducting fraction ratio is about 0.1\% in K-doped terphenyl samples, so the superconducting phase may be a certain structural phases among many possible structures.
As for which structure accounting for superconductivity, it is difficult to identify it.

We compare K$_x$PTP with K-doped condensed aromatics, such as picene, phenanthrene. The obvious distinction is that K positions in K$_x$PTP are close to the single C-C bonds instead of the phenyl rings.
The formation energy of K$_2$-F phase is -0.43 eV per K dopant, greater than the value of about -0.32 eV and 0.33 eV in K$_2$phenanthrene and K$_2$picene, \cite{Yan2016} which indicate that the synthesis of K-doped PTP is more easy than K-doped picene and phenanthrene.
The transition temperatures of superconductivity in K-doped picene, phenanthrene and other condensed aromatic compounds are in the range of 5 Kelvin $\sim$  18 Kelvin, much less than 43 Kelvin or greater in K-doped PTP.
This position difference of K dopant may be the vital information for explain such high temperature in K-doped PTP.

The above calculations and data analysis are based on the assumption that the $p$-terphenyl molecule keeps intact in the synthesis process of K$_x$PTP. But in fact, when the terphenyl and K metal are mixed and heated in experiment, the potassium hydride (KH) will be generated alongside of the $p$-terphenyl molecule damage. The existence of KH in samples can be deduced from the XRD spectra, and (011), (112), (314) peaks in Fig. 1 (b) in Ref. \citenum{Liu2017} are the typical XRD diffraction peaks of KH. To protect the molecule not being destroyed in the process of K$_x$PTP synthesis, KH may be the better reactant than K metal, as done in Ref \citenum{Romero2017}.

\begin{acknowledgments}
We thank Zhong-Yi Lu, Xiao-Jia Chen, Guo-Hua Zhong and Hai-Qing Lin for valuable discussions. This work was supported by the National Natural Science Foundation of China under Grants
Nos. 11474004, 11404383, 11674087, 11704024.
\end{acknowledgments}

\bibliography{KPHA}

\end{document}